\documentclass[prb,preprint,showpacs,preprintnumbers,amsmath,amssymb]{revtex4}

\usepackage{graphicx}
\usepackage{dcolumn}
\usepackage{bm}
\usepackage{color}

\def\be{\begin{equation}}       \def\ee{\end{equation}}
\def\bea{\begin{eqnarray}}      \def\eea{\end{eqnarray}}
\def\half{\frac{1}{2}}
\def\dag{\dagger}
\def\non{\nonumber}

\begin{document}

\title{Effects of ring exchange interaction on the N\'{e}el phase of two-dimensional, spatially anisotropic, frustrated Heisenberg quantum antiferromagnet}

\author{Kingshuk Majumdar}
\email{majumdak@gvsu.edu}
\author{Douglas Furton}
\email{furtond@gvsu.edu}
\affiliation{Department of Physics, Grand Valley State University, Allendale, 
Michigan 49401, USA}

\author{G\"{o}tz S. Uhrig}
\affiliation{Lehrstuhl f\"{u}r Theoretische Physik I, Technische Universit\"{a}t Dortmund,
Otto-Hahn Stra{\ss}e 4, 44221 Dortmund, Germany}
\email{goetz.uhrig@tu-dortmund.de}

\date{\today}

\begin{abstract}
\label{abstract}
Higher order quantum effects on the magnetic phase diagram induced by four-spin ring exchange
on plaquettes are investigated for a two-dimensional quantum antiferromagnet with $S=1/2$.
Spatial anisotropy and frustration are allowed for.
Using a perturbative spin-wave expansion up to second order in $1/S$ we obtain the spin-wave energy dispersion, sublattice magnetization, and the magnetic phase diagram. 
We find that for substantial four-spin ring exchange the quantum fluctuations are stronger 
than in the standard Heisenberg model.
A moderate amount of  four-spin ring exchange couplings stabilizes the ordered
antiferromagnetic N\'eel state while a large amount renders it unstable.
Comparison with inelastic neutron scattering data points toward a moderate ring 
exchange coupling of 27\% to 29\% of the nearest-neighbor exchange coupling.
\end{abstract}


\maketitle

\section{\label{sec:Intro}Introduction}

Despite the intense experimental and theoretical activities to understand the origin of 
high temperature superconductivity in layered oxide high-temperature superconductors, the underlying microscopic mechanism is still
incomplete.\cite{coldea01,headi10,ronnow01,chris04,chris07,bombardi04,melzi00,melzi01,carretta02} 
Very recently the crucial role of magnetic excitations in these compounds has been supported
by their observation in the whole Brillouin zone up to high energies and high levels of doping.\cite{letac11}

\begin{figure}[httb]
\centering
\includegraphics[width=2.0in,clip]{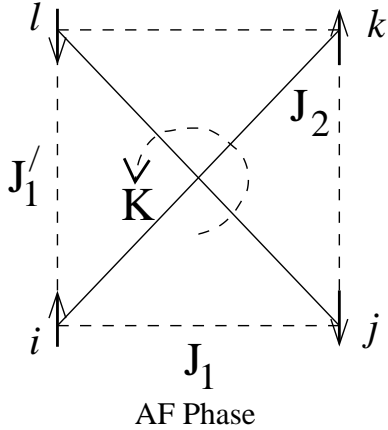}
\caption{\label{fig:GSstates} Classical antiferromagnetic ground state (N\'eel state) and the various
couplings: $J_1$, $J_1^\prime$ are nearest neighbor interactions along the row and column directions respectively, $J_2$ is the next-nearest neighbor interaction along the diagonals, and $K$ is the cyclic four-spin ring exchange. All couplings are assumed to be antiferromagnetic, i.e., 
$J_1, J_1^\prime, J_2, K >0$.}
\end{figure}

The conventional route to theoretically investigate the magnetic properties of these undoped compounds is the two-dimensional (2D) antiferromagnetic (AF) spin-$1/2$ Heisenberg model with nearest neighbor (NN) AF coupling $J_1$ and next-nearest neighbor (NNN) antiferromagnetic coupling $J_2$.\cite{diep}
For concreteness, we give the studied Heisenberg Hamiltonian for a $S=1/2$ antiferromagnet on a square lattice 
\bea
H &=& \half J_1 \sum_{i}{\bf S}_{i} \cdot {\bf S}_{i+\delta_x}
+ \half J_1^\prime \sum_{i}{\bf S}_{i} \cdot {\bf S}_{i+\delta_y} 
+ \half J_2 \sum_{i}{\bf S}_{i} \cdot {\bf S}_{i+\delta_x + \delta_y} \non \\ 
&+& 2K\sum_{i}
\Big[({\bf S}_{i} \cdot {\bf S}_{i+\delta_x})({\bf S}_{i+\delta_y} \cdot {\bf S}_{i+\delta_x+\delta_y})
+ ({\bf S}_{i} \cdot {\bf S}_{i+\delta_y})({\bf S}_{i+\delta_x} \cdot {\bf S}_{i+\delta_x+\delta_y}) \non \\ 
&-&({\bf S}_{i} \cdot {\bf S}_{i+\delta_x+\delta_y})({\bf S}_{i+\delta_y} \cdot {\bf S}_{i+\delta_x})
\Big].
\label{hamiltonian}
\eea
We consider four different exchange couplings: $J_1$ for nearest neighbors (NN) along the rows,
$J_1^\prime$ for NN along the columns, $J_2$ for the next nearest neighbors (NNN) along the diagonals,
and finally the four-spin ring exchange interaction $K$.
All interactions are assumed to be antiferromagnetic, i.e., $J_1,J_1^\prime,J_2, K >0$.
Here $i$ runs over $N$ lattice sites and $\delta_x, \delta_y$ are unit vectors in both directions.
In the present work, we study the parameter region where 
the ground state is of  Ne\'{e}l type as shown in Fig.\ \ref{fig:GSstates}. 
We take $J_1$ as the fundamental energy scale so that the ground state and its properties depend on
the dimensionless ratio $\eta:=J_2/J_1$ parametrizing the degree of frustration,
the ratio $\zeta:=J'_1/J_1$ parametrizing the degree of spatial anisotropy, and the ratio
$\mu=KS^2/J_1$ parametrizing the relative strength of the four-spin ring exchange.
Note that the full cyclic permutation around a plaquette comprises also
two-point couplings along the plaquette edges and along the diagonals.\cite{brehmer99}
But they do not need to be considered separately because they are incorporated in
$J_1$, $J_1'$, and $J_2$.

Experimentally the ground state phase diagram of these frustrated spin systems can be explored from high values to low values of $\eta$ by applying high pressures. For example, X-ray diffraction measurements on
 Li$_2$VOSiO$_4$ show that the value of $\eta$ decreases by about 40\% with increase in pressure from
  zero to 7.6 GPa.~\cite{pavarini08} 

Theoretically, evidence for sizable four-spin ring exchange $K$ in high-temperature
superconductors \cite{roger89,schmi90b,chubukov92} was found soon after the discovery of high-temperature superconductivity.\cite{bedno86} Such exchange processes turned out to be
the dominant subleading correction to the NN Heisenberg Hamiltonian if
it is derived from a three-band Hubbard model \cite{mizun99,mulle02,calza03} 
or from a single-band Hubbard model \cite{takah77,macdo88,reisc04,hamer10}.
Experimental evidence for ring exchange stems from the analysis of infrared absorption
\cite{loren99c}, of Raman response \cite{katanin03,schmi05a}, and of inelastic neutron
scattering \cite{coldea01,headi10,katanin02,uhrig04a}. The results indicate that the
ring exchange coupling reaches between $x_\text{ring}=2K/J_1=0.2$ and 0.25 relative
to the NN coupling. Note that for $S=1/�2$ one has $x_\text{ring}=8\mu$.
These findings and the quantitative estimates
 are strongly supported by the analysis of two-leg spin ladder systems
such as  Sr$_{14}$Cu$_{24}$O$_{41}$, Ca$_8$La$_6$Cu$_{24}$O$_{41}$, and (Ca, La)$_{14}$Cu$_{24}$O$_{41}$.\cite{brehmer99,matsuda00,nunner02,schmi05a,schmi05b,notbo07} 

The recent discovery of superconductivity in the class of iron pnictide has ushered a renewed interest 
in this exciting field.\cite{kamihara08} The parent phases of these materials have been found to be
metallic, but with columnar AF order.\cite{cruz08,klaus08,Dong08}. 
Since the superconductivity appears
in immediate proximity of the magnetically ordered phase, it is evident that the
magnetic excitations play an important role.\cite{zhao08,diallo09,zhao09} 
Neglecting the metallicity of the parent phases
the magnetic excitations can be described by frustrated two-dimensional Heisenberg Hamiltonians
with $S>1/2$ \cite{si08,yao08,uhrig09,singh09} although the three-dimensionality
 cannot be neglected \cite{apple10,yao10,holt11,majum11b}. Ab initio calculations seem to indicate
 a strong spatial anisotropy $\zeta \approx0$ of the NN couplings \cite{han09} fitting to the experimental findings.\cite{zhao08,diallo09,zhao09} But the weak structural distortion does not
 explain this strong anisotropy. So either orbital order \cite{kruge09,singh09a} 
 or higher order magnetic
 exchange such as NN biquadratic coupling \cite{yares09a,wyso11,stane11,yu12} 
 may effectively explain the anisotropy.
 
Another class of magnetic materials described by the Hamiltonian in Eq.\ \eqref{hamiltonian}
are vanadium phosphates. Extensive band structure calculations~\cite{tsirlin09} yielded
four different exchange couplings: J$_1$ and J$_1^\prime$ between the NN and J$_2$ and J$_2^\prime$ between NNN in the compounds 
 Pb$_2$VO(PO$_4$)$_2$, SrZnVO(PO$_4$)$_2$, BaZnVO(PO$_4$)$_2$, and BaCdVO(PO$_4$)$_2$. For example $\zeta \approx 0.7$ and $J_2^\prime/J_2 \approx 0.4$ were obtained for SrZnVO(PO$_4$)$_2$. Also
 the compound (NO)Cu(NO$_3$)$_3$ possibly realizes the 
 $J_1$-$J_1^\prime$-$J_2$ model.~\cite{volkova10}

The above examples corroborate the relevance of the model \eqref{hamiltonian}.

It is now well known that at low temperatures the spin-$1/2$ antiferromagnetic $J_1$-$J_2$ model on a
square lattice  exhibits new types of magnetic order and novel quantum phases.\cite{diep} 
For $J_2=0$ and $K=0$ the ground state is N\'eel ordered at zero temperature. 
Addition of next-nearest neighbor (NNN) interactions induces
a strong frustration and breaks the N\'eel order at a quantum critical point $J_2/J_1 \approx 0.4$ as
found by $1/S$ expansions \cite{igar92,igar05,majumdar10}, series expansion about
the Ising limit \cite{oitma96c}, and the coupled-cluster approach \cite{bisho08a}.
We stress that the precise nature of the phase beyond the N\'eel phase is still
intensely debated \cite{singh99b,sushk01,sirke06}.

A generalization of the frustrated $J_1$-$J_2$ model is the $J_1$-$J_1^\prime$-$J_2$ model where
$\zeta=J_1^\prime/J_1$ is the directional anisotropy parameter.\cite{nersesyan03,igar05,majumdar10}
Recently, the role of directional anisotropy on the magnetic phase diagram has been investigated in detail using a spin-wave expansion.\cite{majumdar10} 

The next generalization consists in the inclusion of the four-spin ring exchange interaction $K$
which is the next important coupling after the NN exchange coupling. Using linear spin-wave theory its effects on the magnetic properties of the $J_1$-$J_2$-$K$ model were studied in
Ref.~\onlinecite{chubukov92} where a quasiclassical phase diagram in ${\cal O}((1/S)^0)$ was obtained. 
In Ref.~\onlinecite{katanin02} corrections to the spin-wave spectrum to first order in $1/S$ 
were studied for finite $K$ using self-consistent spin-wave theory. The 
self-consistent spin-wave theory is a mean-field approach which captures only a part of the second-order effects ${\cal O}((1/S)^2)$ in the phase diagram. In particular, it does not take virtual excitations
of two and four magnons into account. To consider them a perturbative spin-wave expansion up to $1/S^2$
 is needed. That is the goal of the present work.

In the present paper we investigate the higher-order quantum corrections due to the presence of 
plaquette four-spin ring interactions on the antiferromagnetic phase diagram of the 
$J_1$-$J_1^\prime$-$J_2$-$K$ Heisenberg model on a square lattice, cf.\ 
Eq.\ \eqref{hamiltonian}. Our calculations use the Dyson-Maleev spin representation
which facilitates the calculation significantly compared to the Holstein-Primakov representation.
The concomitant formalism is presented in the next section. 
Results for the spin-wave energies and the magnetizations of the system are presented and discussed in Section~\ref{sec:results}. A quantitative comparison with experimental data is also included.
Section~\ref{sec:conclusions} contains a brief summary of our results. 

\section{\label{sec:model} Formalism}

Quantum fluctuations play a significant role in the magnetic phase diagram of the system at zero
temperature. We will investigate the role of quantum fluctuations on the stability of the N\'{e}el
phase. We first express the fluctuations around the classical antiferromagnetic ground state in 
terms of the boson operators using the Dyson-Maleev representation. The quadratic term in boson
operators corresponds to the linear spin-wave theory, whereas the higher-order terms 
represent spin-wave interactions and virtual processes. 
We keep terms up to second order in $1/S$. In the
next step we calculate the renormalized magnon Green's functions and self-energies. Finally,
we calculate the magnon energy dispersion and the sublattice magnetization up  to and including
terms of order $1/S^2$.

For the N\'eel ordered phase NN couplings interact between the A and B sublattices while
NNN couplings link A and A  sites or B and B sites, respectively. 
The Hamiltonian in Eq.~\eqref{hamiltonian} takes the form
\bea
H &=&  J_1 \sum_{i}{\bf S}_{i}^{\rm A} \cdot {\bf S}_{j}^{\rm B}
+J_1^\prime \sum_{i}{\bf S}_{i}^{\rm A} \cdot {\bf S}_{\ell}^{\rm B} 
+ \half J_2 \sum_{i}\Big[ {\bf S}_{i}^{\rm A} \cdot {\bf S}_{k}^{\rm A}
+ {\bf S}_{j}^{\rm B} \cdot {\bf S}_{\ell}^{\rm B}\Big] \non \\
&+& 2K\sum_{i}
\Big[({\bf S}^{\rm A}_{i} \cdot {\bf S}^{\rm B}_{j})({\bf S}^{\rm A}_{k} \cdot {\bf S}^{\rm B}_{\ell})
+({\bf S}^{\rm B}_{j} \cdot {\bf S}^{\rm A}_{k})({\bf S}^{\rm B}_{\ell} \cdot {\bf S}^{\rm A}_{i}) 
-({\bf S}^{\rm A}_{i} \cdot {\bf S}^{\rm A}_{k})({\bf S}^{\rm B}_{\ell} \cdot {\bf S}^{\rm B}_{j})
\Big],
\label{ham-AF}
\eea
where $j=i+\delta_x,\;k=i+\delta_x+\delta_y,\;\ell=i+\delta_y$ as shown in Fig.\ \ref{fig:GSstates}.
Beside the directional anisotropy parameter $\zeta = J_1^\prime/J_1$, the magnetic frustration between the NN and NNN spins $\eta=J_2/J_1$, and the cyclic four-spin exchange interaction term $\mu=KS^2/J_1$
we use $z=2$ for the coordination number. 
This spin Hamiltonian is mapped onto an equivalent Hamiltonian
of interacting bosons by expressing the spin operators in terms of bosonic creation 
and annihilation operators 
$a^\dag, a$ for ``up'' sites on sublattice A and $b^\dag, b$ for ``down'' sites on  sublattice 
B using the  Dyson-Maleev representation
\begin{subequations}
 \label{dyson} 
\begin{eqnarray}
S_{Ai}^+ &=& \sqrt{2S}\Big[a_i-  \frac {a_i^\dag a_i a_i}{(2S)}\Big],\;
S_{Ai}^-= \sqrt{2S}a_i^\dag,\;
S_{Ai}^z = S-a^\dag_ia_i,  \\ 
S_{Bj}^+ &=& \sqrt{2S} \Big[b_j^\dag- \frac {b_j^\dag b_j^\dag b_j}{(2S)} \Big],\;
S_{Bj}^- = \sqrt{2S} b_j,\;
S_{Bj}^z = -S+b^\dag_jb_j. 
\end{eqnarray}
\end{subequations}

Substituting Eqs.~\eqref{dyson} into \eqref{ham-AF} 
we expand the Hamiltonian perturbatively in powers of $1/S$ as
\be
H = H_{-1}+H_0+H_1+H_2 + \cdots,
\ee 
where $H_m$ is of order $1/S^{m-1}$. Note that $H_{-1}$ is just a number 
representing the classical energy. We do not discuss it further because
it is irrelevant for the quantum fluctuations. Hence the $1/S$ expansion will
be performed around the unperturbed Hamiltonian $H_0$ which is 
the zeroth order Hamiltonian in this sense. Relative to $H_0$ the terms
$H_1$ and $H_2$ are first and second order terms, respectively.

Next the real space Hamiltonian is Fourier transformed to momentum space. Then
we diagonalize the quadratic part $H_0$ by transforming the 
operators $a_{\bf k}$ and $b_{\bf k}$ to magnon operators 
$\alpha_{\bf k}$ and $\beta_{\bf k}$ using the usual Bogoliubov (BG) transformations
\be
a^\dag_{\bf k} =l_{\bf k} \alpha_{\bf k}^\dag + m_{\bf k}\beta_{-{\bf k}},\;\;\;
b_{-\bf k} =m_{\bf k} \alpha_{\bf k}^\dag + l_{\bf k}\beta_{-{\bf k}}.
\ee
The coefficients $l_{\bf k}$ and $m_{\bf k}$ are defined as
\be
l_{\bf k} = \Big[\frac {1+\epsilon_{\bf k}}{2\epsilon_{\bf k}} \Big]^{1/2},\;\;
m_{\bf k} = -{\rm sgn}(\gamma_{\bf k})\Big[\frac {1-\epsilon_{\bf k}}
{2\epsilon_{\bf k}} \Big]^{1/2}\equiv -x_{\bf k}l_{\bf k},\;\;
x_{\bf k} = {\rm sgn}(\gamma_{\bf k})\Big[\frac {1-\epsilon_{\bf k}}
{1+\epsilon_{\bf k}} \Big]^{1/2},
\ee
with $\gamma_{kx}=\cos (k_x),\;\gamma_{ky}=\cos (k_y)$ and 
\begin{subequations}
 \label{defs-AF}
\bea
\epsilon_{\bf k} &=& (1-\gamma_{\bf k}^2)^{1/2}, \\
\gamma_{\bf k} &=& \frac {\gamma_{1{\bf k}}}{\kappa_{\bf k}}, \\
\gamma_{1{\bf k}}&=&\frac {(1-4\mu)\gamma_{kx}+(\zeta -4\mu)\gamma_{ky}}{1+\zeta-8\mu}, \\
\gamma_{2{\bf k}} &=& \gamma_{kx}\gamma_{ky},\\
\kappa_{\bf k} &=& 1- \frac {2(\eta-2\mu)}{1+\zeta-8\mu} (1-\gamma_{2{\bf k}}).
\eea
\end{subequations}
The function ${\rm sgn} (\gamma_{\bf k})$ keeps track of the sign
of $\gamma_{\bf k}$ in the first Brillouin zone (BZ). 
After these transformations, the quadratic part of the Hamiltonian takes the form
\be
H_0 = J_1Sz(1+\zeta-8\mu)\sum_{\bf k} \kappa_{\bf k}\left(\epsilon_{\bf k}-1\right)
+J_1Sz(1+\zeta-8\mu)\sum_{\bf k}\kappa_{\bf k} \epsilon_{\bf k}
\left( \alpha^\dag_{\bf k}\alpha_{\bf k}+\beta^\dag_{\bf k}\beta_{\bf k}\right).
\label{H0term}
\ee
The first term is the quantum zero-point energy and the second term represents the 
excitation energy of the magnons within linear spin-wave theory (LSWT).\cite{chubukov92}

The part $H_1$ comprises $1/S$ contribution to the Hamiltonian. We follow the same
procedure as described above. The resulting expression after
transforming the bosonic operators to  magnon operators is
\bea
H_1 &=& \frac {J_1Sz(1+\zeta-8\mu)}{2S}\sum_{\bf k}
\Big[ A_{\bf k}\left(\alpha^\dag_{\bf k}\alpha_{\bf k}+
\beta^\dag_{\bf k}\beta_{\bf k}\right) 
+ B_{\bf k}\left(\alpha^\dag_{\bf k}\beta_{-\bf k}^\dag+
\beta_{-\bf k}\alpha_{\bf k}\right)\Big] \non \\
&-& \frac {J_1Sz(1+\zeta-8\mu)}{2SN}\sum_{1234}
\delta_{\bf G}(1+2-3-4)l_1l_2l_3l_4
\Big[V_{12;34}^{(1)} \alpha_1^\dag \alpha_2^\dag \alpha_3 \alpha_4 
+2V_{12;34}^{(2)}\alpha_1^\dag \beta_{-2}\alpha_3 \alpha_4 \non \\
&+& 2V_{12;34}^{(3)}\alpha_1^\dag \alpha_2^\dag \beta_{-3}^\dag \alpha_{4}
+4V_{12;34}^{(4)}\alpha_1^\dag \alpha_3 \beta_{-4}^\dag \beta_{-2}
+2V_{12;34}^{(5)}\beta_{-4}^\dag \alpha_3 \beta_{-2} \beta_{-1}
+2V_{12;34}^{(6)}\beta_{-4}^\dag \beta_{-3}^\dag \alpha_{2}^\dag \beta_{-1} \non \\
&+&V_{12;34}^{(7)}\alpha_1^\dag \alpha_2^\dag \beta_{-3}^\dag \beta_{-4}^\dag
+V_{12;34}^{(8)}\beta_{-1} \beta_{-2}\alpha_3 \alpha_4
+V_{12;34}^{(9)}\beta_{-4}^\dag \beta_{-3}^\dag \beta_{-2} \beta_{-1}\Big].
\label{H1term}
\eea 
In the above equation momenta ${\bf k}_1, {\bf k}_2, {\bf k}_3, {\bf k}_4$ are abbreviated
as 1, 2, 3, and 4. The first term in Eq.~\eqref{H1term} is obtained by 
normal ordering the products of four boson operators with respect to 
creation and annihilation in the magnon operators, i.e., magnon creation operators
appear always to the left of magnon annihilation operators.
The coefficients $A_{\bf k}$ and $B_{\bf k}$ read
\begin{subequations}
\bea
A_{\bf k}&=& A_1 \frac 1{\kappa_{\bf k}\epsilon_{\bf k}}\Big[\kappa_{\bf k}
-\gamma_{1{\bf k}}^2\Big] + A_2 \frac 1{\epsilon_{\bf k}}
\Big[1-\gamma_{2{\bf k}}\Big]+A_3\frac 1{\epsilon_{\bf k}}\Big[(1+\gamma_{2{\bf k}})-\gamma_{\bf k}(\gamma_{kx}+\gamma_{ky})\Big], \\
B_{\bf k} &=& B_1 \frac {1}{\kappa_{\bf k}\epsilon_{\bf k}}
\gamma_{1{\bf k}}\Big[1-\gamma_{2{\bf k}}\Big]+A_3\frac 1{\epsilon_{\bf k}}\Big[(\gamma_{kx}+\gamma_{ky})-\gamma_{\bf k}
(1+\gamma_{2{\bf k}})\Big],
\eea
\end{subequations}
where the shorthands
\begin{subequations}
\bea
A_1 &=& \Big(\frac 2{N}\Big) \sum_{\bf p} \frac 1{\epsilon_{\bf p}}
\Big[\gamma_{\bf p}\gamma_{1{\bf p}}+\epsilon_{\bf p}-1\Big], \\
A_2 &=& \frac {2(\eta-4\mu)}{1+\zeta-8\mu}\Big(\frac 2{N}\Big) \sum_{\bf p} 
\frac 1{\epsilon_{\bf p}}\Big[1-\epsilon_{\bf p}-\gamma_{2{\bf p}}\Big], \\
A_3 &=& \frac {4\mu}{1+\zeta-8\mu}\Big(\frac 2{N}\Big) \sum_{\bf p} 
\frac 2{\epsilon_{\bf p}}\Big[1-\epsilon_{\bf p}+\gamma_{2{\bf p}}-\gamma_{\bf p}(\gamma_{px}+\gamma_{py})\Big], \\
B_1 &=& \frac {2(\eta-2\mu)}{1+\zeta-8\mu}\Big(\frac 2{N}\Big) \sum_{\bf p} 
\frac 1{\epsilon_{\bf p}}\Big[\gamma_{2{\bf p}}-
\gamma_{\bf p}\gamma_{1{\bf p}}\Big]
\eea
\end{subequations}
are used.

The second term in Eq.~\eqref{H1term} represents scattering between spin-waves where the delta
function $\delta_{\bf G}(1+2-3-4)$ ensures that the momentum is conserved within a 
reciprocal lattice vector ${\bf G}$. 
Explicit forms of the vertex factors $V_{1234}^{i=2,3,5,7,8}$ 
are given in Appendix~\ref{VertexAF}.

The second order term, $H_2$ is composed of six-boson operators and is only present when $\mu \neq 0$. Before the Fourier and BG transformations $H_2$ is of the following form
\bea
H_2 &=& -\frac {8\mu S}{(2S)^2}\sum_{i} \Big[ 
(a_i^\dag a_i+b_j^\dag b_j+a_ib_j+a^\dag_i b_j^\dag)
(a_k^\dag a_k b_{\ell}^\dag b_{\ell}+\frac 1{2} a_k^\dag a_k a_k b_{\ell}+\frac 1{2}a_k^\dag b_{\ell}^\dag b_{\ell}^\dag b_{\ell}) \non \\
&+& (a_k^\dag a_k+b_j^\dag b_j+a_kb_j+a^\dag_k b_j^\dag)
(a_i^\dag a_i b_{\ell}^\dag b_{\ell}+\frac 1{2} a_i^\dag a_i a_i b_{\ell}+\frac 1{2}a_i^\dag b_{\ell}^\dag b_{\ell}^\dag b_{\ell}) \non \\
&+& (a_i^\dag a_i b_{j}^\dag b_{j}+\frac 1{2} a_i^\dag a_i a_i b_{j}+\frac 1{2}a_i^\dag b_{j}^\dag b_{j}^\dag b_{j})(a_k^\dag a_k+b_{\ell}^\dag b_{\ell}+a_kb_{\ell}+a^\dag_k b_{\ell}^\dag) \non \\
&+& (a_k^\dag a_k b_{j}^\dag b_{j}+\frac 1{2} a_k^\dag a_k a_k b_{j}+\frac 1{2}a_k^\dag b_{j}^\dag b_{j}^\dag b_{j})(a_i^\dag a_i+b_{\ell}^\dag b_{\ell}+a_i b_{\ell}+a^\dag_i b_{\ell}^\dag) \non \\
&-& (a_i^\dag a_i+a_k^\dag a_k-a_i a_k^\dag-a^\dag_i a_k)
(b_j^\dag b_j b_{\ell}^\dag b_{\ell}-\frac 1{2} b_j^\dag b_j^\dag b_j b_{\ell}-\frac 1{2}b_j b_{\ell}^\dag b_{\ell}^\dag b_{\ell}) \non \\
&-& (a_i^\dag a_i a_{k}^\dag a_{k}-\frac 1{2} a_i^\dag a_ia_i a_k^\dag-\frac 1{2}a_i^\dag a_k^\dag a_k a_k)(b_j^\dag b_j+b_{\ell}^\dag b_{\ell}-b_j b_{\ell}^\dag-b^\dag_j b_{\ell})\Big].
\eea
After Fourier and BG transformations to magnon operators 
$\alpha_{\bf k},\beta_{\bf k}$ the Hamiltonian
in normal-ordered form reduces to
\be
H_2= -\frac {4\mu zS}{(2S)^2} \sum_{\bf k} 
\Big[ {\cal C}_{1{\bf k}}\left(\alpha^\dag_{\bf k}\alpha_{\bf k}+\beta^\dag_{\bf k}\beta_{\bf k}
\right)+{\cal C}_{2{\bf k}}\left(\alpha^\dag_{\bf k}\beta_{-\bf k}^\dag+
\beta_{-\bf k}\alpha_{\bf k}\right)+...\Big].
\label{H2term}
\ee
The dotted terms contribute only to higher than second order corrections and are thus omitted in 
our calculations.  
The coefficients $C_{1{\bf k}}$ and $C_{2{\bf k}}$ are given in Appendix~\ref{Ck}.

The quasiparticle energy ${\tilde E_{\bf k}^{\rm AF}}$ for magnon excitations, measured 
in units of $J_1Sz(1+\zeta-8\mu)$ up to second order in $1/S$ is given as
\be
{\tilde E_{\bf k}^{\rm AF}} = E_{\bf k} + \frac 1{(2S)} A_{\bf k}+
\frac 1{(2S)^2}\Big[\Sigma^{(2)}_{\alpha \alpha}({\bf k},E_{\bf k})-
\frac {B_{\bf k}^2}{2E_{\bf k}} \Big].
\label{energyEk-AF}
\ee
Expressions for the magnon Green's functions and self-energies are given in Appendix~\ref{Greensfunction}.
The dynamic contributions to the second order self-energies $\Sigma^{(2)}$ are second
order in the vertex factors $V^{(j)}$. These are the contributions which are missed
by self-consistent spin-wave theory.

The sublattice magnetization $M_{\rm AF}$ for the A sublattice can be expressed as 
\be
M_{\rm AF} = S-\langle a^\dag_i a_i \rangle = S-\Delta S + \frac {M_1}{(2S)}+\frac {M_2}{(2S)^2},
\label{Mag-AF}
\ee
where
\begin{subequations}
\bea
\Delta S &=& \frac 2{N} \sum_{\bf k} \frac 1{2\epsilon_{\bf k}}-\half, 
\label{MagAF-LSWT} \\
M_1 &=& \frac 2{N} \sum_{\bf k}\frac {l_{\bf k}m_{\bf k}B_{\bf k}}{E_{\bf k}},  
\label{M1-AF} \\
M_2 &=& \frac 2{N} \sum_{\bf k} \Big\{ -(l_{\bf k}^2+m_{\bf k}^2)\frac {B_{\bf k}^2}{4E_{\bf k}^2}
+ \frac {l_{\bf k}m_{\bf k}}{E_{\bf k}}\Sigma^{(2)}_{\alpha \beta}({\bf k},-E_{\bf k}) \non \\
&-& \Big(\frac 2{N} \Big)^2 \sum_{\bf pq} 2l_{\bf k}^2l_{\bf p}^2l_{\bf q}^2l_{\bf k+p-q}^2 
\Big[\frac {(l_{\bf k}^2+m_{\bf k}^2)V^{(7)}_{\bf k,p,q,[k+p-q]}V^{(8)}_{\bf [k+p-q],q,p,k}}
{(E_{\bf k}+E_{\bf p}+E_{\bf q}+E_{\bf k+p-q})^2} \non \\
&+& \frac {2l_{\bf k}m_{\bf k}V^{(7)}_{\bf k,p,q,[k+p-q]}
V^{(5)}_{\bf [k+p-q],q,p,k}}{E_{\bf k}^2-(E_{\bf p}+E_{\bf q}+E_{\bf k+p-q})^2}. \Big]
\label{M2-AF}
\Big\}
\eea
\end{subequations}
The zeroth-order term $\Delta S$ corresponds to the reduction of magnetization within 
LSWT, $M_1$ term corresponds to the first-order $1/S$ correction, and $M_2$ is the second-order
correction. Again, the parts which are second order in the vertex factors
are not captured by self-consistent spin-wave theory.

\section{\label{sec:results}Results}

\subsubsection{\label{sec: AFphase-energy}Spin-Wave Energy}

We obtain the spin-wave energy 
$2J_1S(1+\zeta-8\mu){\tilde E_{\bf k}^{\rm AF}}$ for $S=1/2$ 
as a function of momenta ($k_x, k_y$) for several values
of $\zeta, \eta$, and $\mu$ by evaluating  Eq.~\eqref{energyEk-AF} in the first BZ. For the numerical summation we divide the first BZ in a mesh of $N_L^2$ points with $N_L=48$ and then the contributions
from all the points are summed up to evaluate the third term in Eq.~\eqref{energyEk-AF}.
In the Dyson-Maleev formalism, no cancellation of divergences occurs so that the 
convergence of the numerical results for $N_L\to\infty$ is very good.
This is a crucial advantage over the use of the Holstein-Primakov representation.
We estimate that the results for $N_L=48$ will not change more than 
in the third digit if $N_L$ is chosen larger.

\begin{figure}[httb]
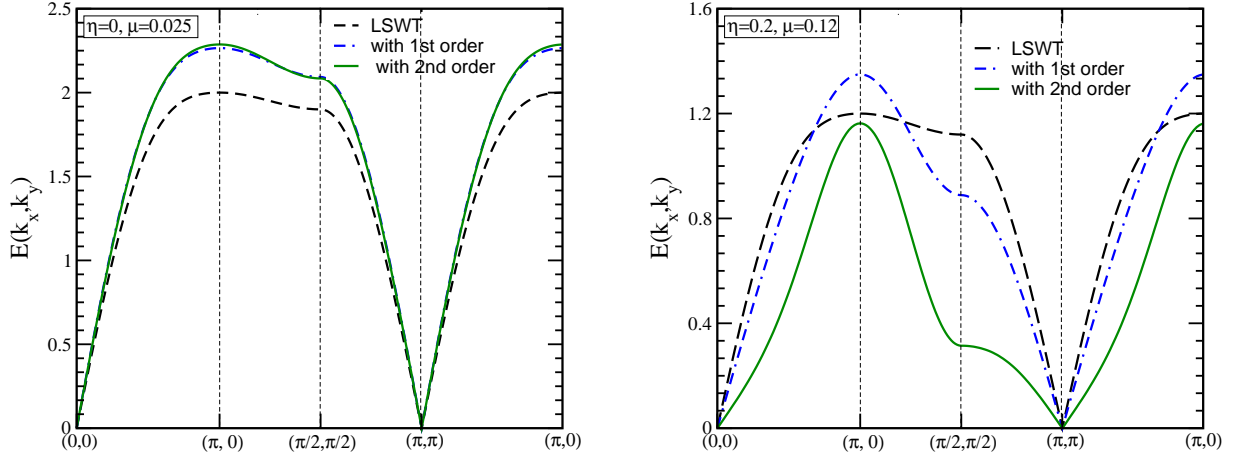

\centering
\includegraphics[width=3.0in,clip]{Compare-energy_eta0-mu025.eps}
\qquad
\includegraphics[width=3.0in,clip]{Compare-energy_eta02-mu12.eps}
\caption{\label{fig:CompareAFenergy} 
(Color online) Spin-wave energy $E_{\bf k}^{\rm AF}/J_1$ obtained from  
LSWT (long-dashed lines), with $1/S$ (dot-dashed lines) and with $1/S^2$ corrections (solid lines)
for the N\'eel-ordered phase. We have chosen spatially isotropic
coupling $\zeta=1$. In the left panel we show the corrections  for relative frustration $\eta=0$ and 
ring exchange  $\mu=0.025$; in the right panel we show them
for frustration $\eta=0.2$ and ring exchange $\mu=0.12$.
In the latter case, the $1/S^2$ terms in the Hamiltonian 
provide significant corrections to both the LSWT and $1/S$ results.}
\end{figure}

Figure \ref{fig:CompareAFenergy} shows a comparison between the results from LSWT (long-dashed lines),
first-order (dot-dashed lines) and second-order corrections (solid lines) to the spin-wave energy spectrum
for isotropic coupling ($\zeta=1$) for two choices of frustration and and ring exchange.
For the moderate value $\mu=0.025$ corresponding to $2K/J_1=0.2$ the $1/S$ correction is substantial
while the $1/S^2$ correction is fairly small. This is very similar to the corrections
for the NN Heisenberg model at $\mu=0$.\cite{igar92,hamer92,zheng93,igar05,majumdar10,syrom10}
The right panel of Fig.\ \ref{fig:CompareAFenergy} tells quite a different story.
For substantial ring exchange the quantum corrections are very large and cannot be ignored.
We point out that this is not due to the frustration alone as can be seen by inspecting
the results for substantial values of $\eta$, but without ring exchange $\mu=0$, in Ref.\ 
\onlinecite{majumdar10}. The $1/S^2$ corrections for $\mu=0$ are as small as they are
for the NN Heisenberg model, in contrast to the result in the right panel of Fig.\
\ref{fig:CompareAFenergy}.

\begin{figure}[httb]
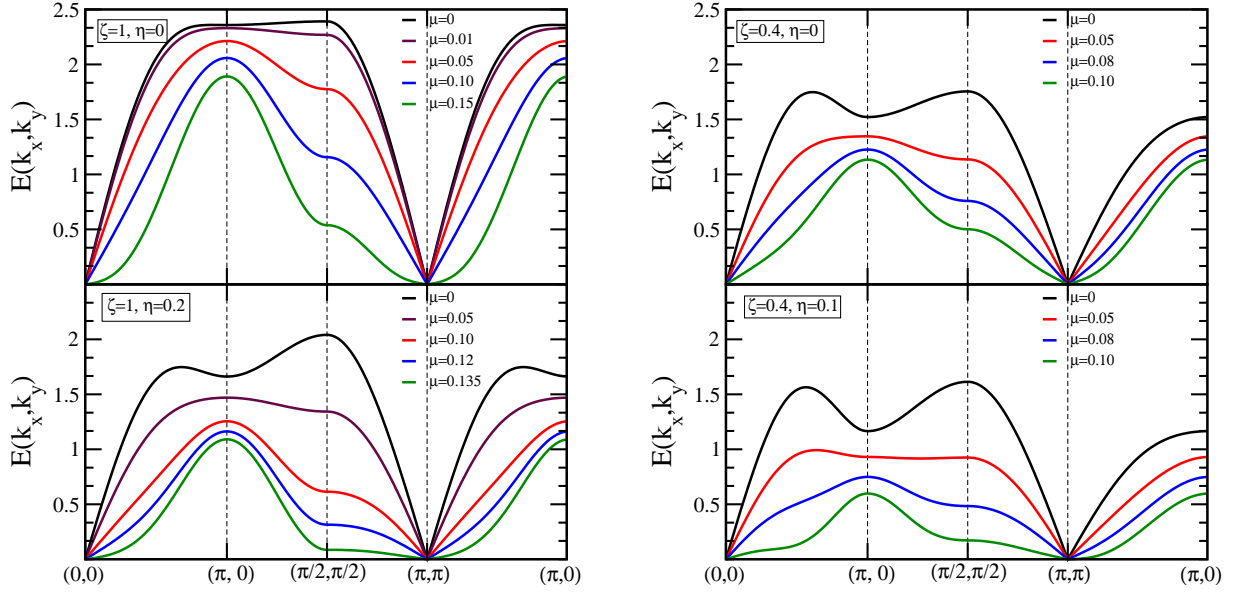

\centering
\includegraphics[width=3.0in,clip]{energy_zeta1.eps}
\qquad 
\includegraphics[width=3.0in,clip]{energy_zeta04.eps}
\caption{(Color online) The effect of $\mu$ on the spin-wave energy $E_{\bf k}^{\rm AF}/J_1$ for the 
N\'eel-ordered phase with $1/S^2$ corrections is shown for two values of $\eta=0, 0.2$ and $\zeta=1$. }
\label{fig:AFenergydisp}
\end{figure}

In the panels of Fig.\ \ref{fig:AFenergydisp} the evolution of the spin-wave energy spectrum including corrections up to second-order for various values of $\zeta, \eta$ and $\mu$ are shown. 
The spin-wave dispersions for the couplings $\zeta=1$ and $\zeta=0.4$ at $\mu=0$ were reported earlier using the Holstein-Primakov representation.\cite{majumdar10} The results from the Dyson-Maleev and
from the Holstein-Primakov representation coincide as it has to be for physically observable 
results of a systematic expansion in a small parameter.

For $\mu=0$ and $\eta=0$, the energy at $(\pi/2, \pi/2)$ is larger than the energy at $(\pi, 0)$,
cf.\ upper left panel in Fig.\ \ref{fig:AFenergydisp}.
This dip of the dispersion at $(\pi, 0)$ has been first computed by high-order
series expansion (HSE) around the Ising limit \cite{singh95,weihong05} and was confirmed by quantum Monte
Carlo calculation (QMC).\cite{sandv01} 
HSE and QMC find that
the dip is about 9\% deep, i.e., $[E((\pi/2, \pi/2))-E((\pi, 0))]/E((\pi/2, \pi/2))\approx 0.09$.
Experimentally, the dip is found to be about 7\% in compounds in which no couplings
beyond $J_1$ are thought to play a role, in reasonable agreement with HSE and QMC.\cite{chris04,chris07} 

In contrast, LSWT and order $1/S$ do not find a dip at all. In order $1/S^2$, it is present but
as small as $1.4$\% and in order $1/S^3$ it takes the value of $3.2$\%.\cite{syrom10}
Thus one must be aware that the data in Fig.\ \ref{fig:AFenergydisp} does not
capture all aspects of the dispersion between $(\pi, 0)$ and $(\pi/2, \pi/2)$. 
But in the remaining BZ the significance
of corrections of third order and higher is rather small and the agreement with the series
expansion results very good.

Having the above minor caveat in mind, we discuss the much stronger influence of
frustration and of ring exchange in the following.
Increasing the value of $\mu$ to positive values
the energy at ($\pi/2,\pi/2$) decreases more strongly than the one
at $(\pi, 0)$, see left panels of Fig.\ \ref{fig:AFenergydisp}. 
Hence, beyond some finite value of four-spin ring exchange 
there is a dip from $(\pi, 0)$ to $(\pi/2, \pi/2)$. This agrees qualitatively with 
experimental findings \cite{coldea01,headi10}, which see a 13\% dip,
and with an analysis based on self-consistent spin-wave theory.\cite{katanin02}
Even larger values of $\mu$ will lead to a complete softening of the magnon mode
at $(\pi/2,\pi/2)$. This indicates a competition between an ordered orthogonal state at modulation 
($\pi/2,\pi/2$) and  the ordered  N\'eel state at ($\pi,\pi$) upon increasing $\mu$.

\begin{figure}[httb]
\centering
\includegraphics[width=3.5in,clip]{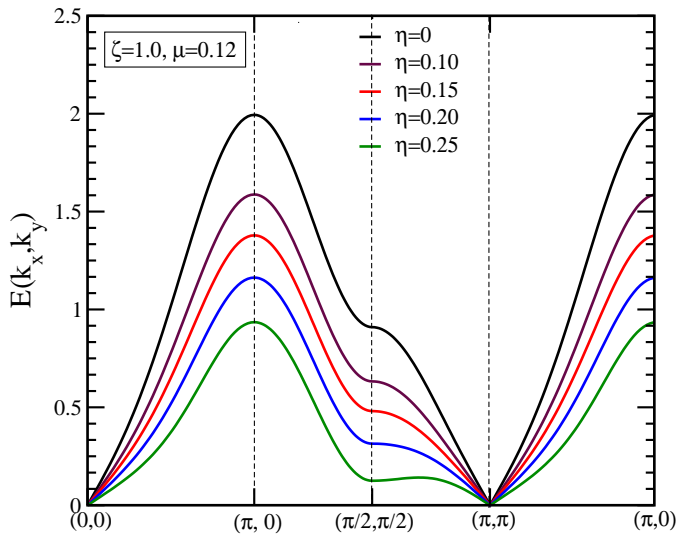}
\caption{\label{fig:energy-eta} 
(Color online) Spin-wave energy $E_{\bf k}^{\rm AF}/J_1$  
including $1/S^2$ corrections for $\zeta=1, \mu=0.12$ for various values of $\eta$.}
\end{figure}

Another important issue is the effect of finite frustration $\eta>0$ which has been 
investigated before without ring exchange.\cite{igar05,majumdar10}
Indeed, finite frustration induces a significant dip at $(\pi,0)$ relative to $(\pi/2,\pi/2)$,
i.e., $E((\pi,0)) < E((\pi/2,\pi/2))$, so that frustration pushes the system
into the opposite direction as does the ring exchange.
But in the presence of \emph{substantial} ring exchange the effect is reversed:
Comparing the upper and lower left panels in Fig.\ \ref{fig:AFenergydisp}
and inspecting Fig.\ \ref{fig:energy-eta} we see that increasing frustration
supports the tendency to soften the mode at $(\pi/2,\pi/2)$ which will eventually
destabilize the N\'eel order.

Spatial anisotropy, see right panel in Fig.\ \ref{fig:AFenergydisp}, does not alter this
picture qualitatively. A strong anisotropy $\zeta < 1$ seems to support the tendency
to mode softening and the concomitant destabilization of the  N\'eel order.

\subsubsection{\label{sec:ins-analysis} Quantitative Analysis of the Inelastic
Neutron Scattering Data}

We use our model to quantitatively analyse the experimental data obtained in
Ref.\ \onlinecite{headi10} by inelastic neutron scattering for La$_2$CuO$_4$. 
We disregard any spatial anisotropy because La$_2$CuO$_4$ is tetragonal
so that we set $\zeta=1$. The experimental data displays a significant
dip at $(\pi/2,\pi/2)$ relative to the energy at  $(\pi,0)$. This points
toward a sizable four-spin ring exchange \cite{coldea01,katanin02}.

\begin{figure}[httb]
\centering
\includegraphics[width=3.5in,clip]{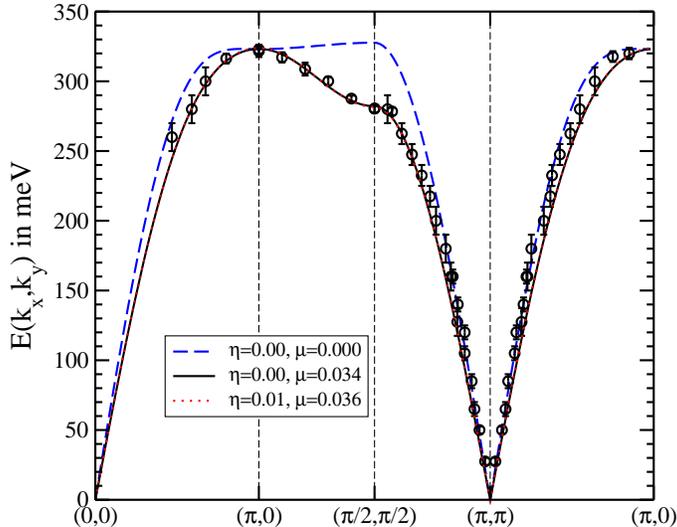}
\caption{\label{fig:ins-analysis} 
(Color online) Comparison of the measured spin-wave energy $E_{\bf k}^{\rm AF}$  
as obtained by inelastic neutron scattering in La$_2$CuO$_4$ with the theoretical
results including $1/S^2$ corrections for the spatially isotropic model ($\zeta=1$)
for $N_L=24$.
For given moderate values $\eta$ of relative frustration a value $\mu$ of the
four-spin ring exchange can be found such that the dispersions match the experimental data.
}
\end{figure}

Our findings are shown in Fig.\ \ref{fig:ins-analysis}. They
strikingly confirm that substantial values of $\mu$ are needed to
explain the observed energy dip at $(\pi/2,\pi/2)$.
For instance, for $\eta=0$ one needs $\mu=0.034$, and $J_1=143$ meV;
for $\eta=0.01$  $\mu=0.036$, and $J_1=146$ meV; 
for $\eta=0.02$  $\mu=0.0375$, and $J_1=148$ meV (not shown).
 Even for $\eta=0.10$ the parameters $\mu=0.046$, and $J_1=174$ meV 
 yield a theoretical dispersion which is indistinguishable 
 from those displayed in Fig.\ \ref{fig:ins-analysis}.
 Note that the agreement of the steeply rising parts of the 
 dispersion is not completely perfect because the theoretical curves
 remain a bit below the experimental data points.
 
 We conclude that from the experimental data for the spin-wave energies the
 relative frustration and the relative ring exchange cannot both be determined 
 independently. Based on the results of systematic derivations of
 extended Heisenberg models for the cuprates starting from microscopic
 Hubbard models \cite{mulle02,reisc04,hamer10} we stick to small values of
 frustration $\eta\approx 0.01$. According to our fits this implies $x_\text{ring}
 =2K/J_1 =8\mu=0.29$. This relative four-spin ring exchange is 
 slightly larger than we would expect from the
 systematic derivations.\cite{mulle02,reisc04,hamer10}
 It is also slightly larger than the value $0.24$ found in
 the analysis by self-consistent spin-wave theory.\cite{katanin02}
 
 On the one hand, the agreement is good in view of the remaining uncertainty 
 in the description of the spin-wave energies at the zone boundary, 
 see our discussion above. On the other hand, a further improved 
 theoretical treatment of spin-waves is desirable.

\subsubsection{\label{sec: AFphase-mag}Sublattice Magnetization and the Phase Diagram}

We calculate the sublattice magnetization $M_{\rm AF}$ from Eq.~\eqref{Mag-AF} by numerically evaluating Eqs.~\eqref{MagAF-LSWT}--\eqref{M2-AF} with $\zeta=1$ and 0.8 and for 
$\mu=0, 0.12$ and 0.22. Especially to obtain the second order correction term $M_2$ we sum
up the values of $N_L^2/4$ points of ${\bf k}$ in a quarter of the 
first BZ and $N_L^2$ points of ${\bf p}$ and ${\bf q}$ in the first BZ,
with $N_L=36$ sites along one axis.

\begin{figure}[httb]
\centering
\includegraphics[width=3.5in,clip]{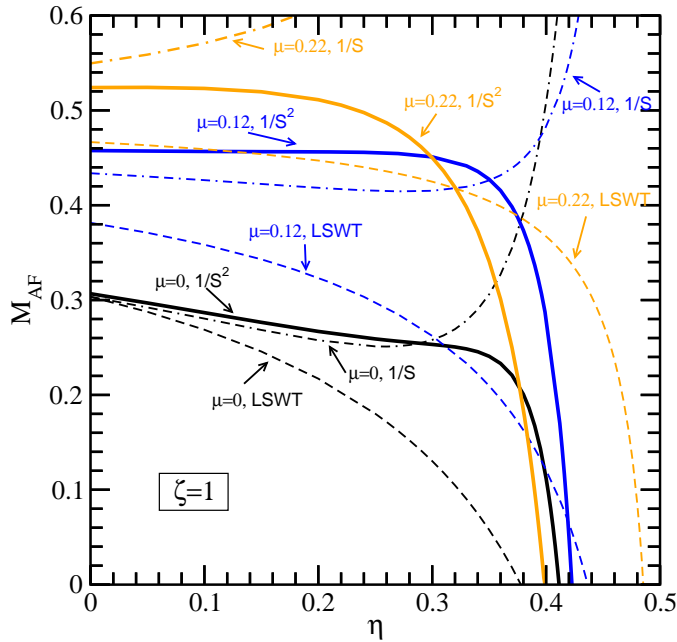}
\caption{(Color online) 
The sublattice magnetization $M_{\rm AF}$ is plotted for $\zeta=1$ and for three different values
of $\mu=0$ (black), 0.12 (blue/dark gray), 0.22 (orange/light gray) 
as a function of the relative magnetic frustration $\eta$. 
For all three cases, results from linear spin-wave theory (dashed lines), with $1/S$ (dot-dashed lines), and with $1/S^2$ corrections (solid lines) are shown.
Magnetization curves with $1/S$ corrections alone diverge in all cases. 
However, $1/S^2$ corrections compensate the 
divergence and the magnetization curves steadily decrease to zero at critical values $\eta_c$ 
We find $\eta_c=0.411$ ($\mu=0$), 0.423 ($\mu=0.12$), and 0.399 ($\mu=0.22$). }
\label{fig:magdia1}
\end{figure}

Figure \ref{fig:magdia1} shows the sublattice magnetization with 
increase in the frustration parameter $\eta=J_2/J_1$ for the isotropic case 
$\zeta = J_1^\prime/J_1 = 1$ for three different values of plaquette ring
exchange coupling $\mu=KS^2/J_1=0, 0.12$, and 0.22. For each case, 
three different curves are plotted: 
The long-dashed lines represent the LSWT prediction,
the dotted lines include the first-order ($1/S$) correction to the LSWT results, and the solid lines 
include corrections up to second-order ($1/S^2$). Upon increasing frustration
the dotted curves of the first-order corrections diverge. 
However, $1/S^2$ corrections ($M_2$) significantly increase
with frustration and stabilize the apparent divergence of the magnetization.
We find that the magnetization with second-order corrections decreases steadily at first with increase in
$\eta$ and then sharply drops to zero at a critical value of $\eta=\eta_c$. 
Assuming that the N\'eel phase loses its stability continuously, $\eta_c$ marks
the quantum critical point at which the AF order is destroyed and the 
system enters into another state characterized by other types of order.
The precise order of the phase transition and the nature of the subsequent phase is still
matter of intense debate.\cite{singh99b,sushk01,sirke06}.

Without four-spin ring exchange, i.e., $\mu=0$, $M_{\rm AF}$ with second-order corrections begins from 0.307 at $\eta=0$ and decreases upon rising frustration till $\eta \approx 0.32$.
Finally it vanishes at $\eta_{c1} \approx 0.411$. For this case, we reproduce the magnetization plot obtained in Ref.~\onlinecite{majumdar10} using a similar perturbative $1/S$ expansion based on the
Holstein-Primakov representation. The LSWT prediction for the critical point is lower at $\approx0.38$.
With increase in the four-spin ring exchange $\mu$ the values of the magnetization at $\eta=0$ increase.
For example, we find $M_{\rm AF}(\eta=0, \mu=0.12) \approx 0.458$ and 
$M_{\rm AF}(\eta=0, \mu=0.22) \approx 0.524$.
These numbers are significantly larger than the predictions from LSWT 
which are 0.381 and 0.466, respectively.
We conclude that without NNN frustration ($\eta=0$) the pure four-spin coupling $\mu$ favors the
N\'eel order. This is in qualitative accord with the observation that the spin gap of the disordered
paramagnetic phase of spin ladders is reduced on increasing four-spin coupling $\mu$.\cite{brehmer99,matsuda00,schmi05b,notbo07}
Thus finite four-spin coupling pushes spin ladders closer to a gapless phase which
is likely to display quasi-long range order with powerlaw correlations.

We observe that first and second order corrections provide significant contributions to the 
entire magnetization curves.  For small $\mu$,  the corrections $M_2$  start from a small positive value
and then switch sign and become negative with increase in $\eta$. However,
for large  $\mu$, say $\mu=0.22$ $M_2$, corrections are negative throughout. 

\begin{figure}[httb]
\centering
\includegraphics[width=3.5in,clip]{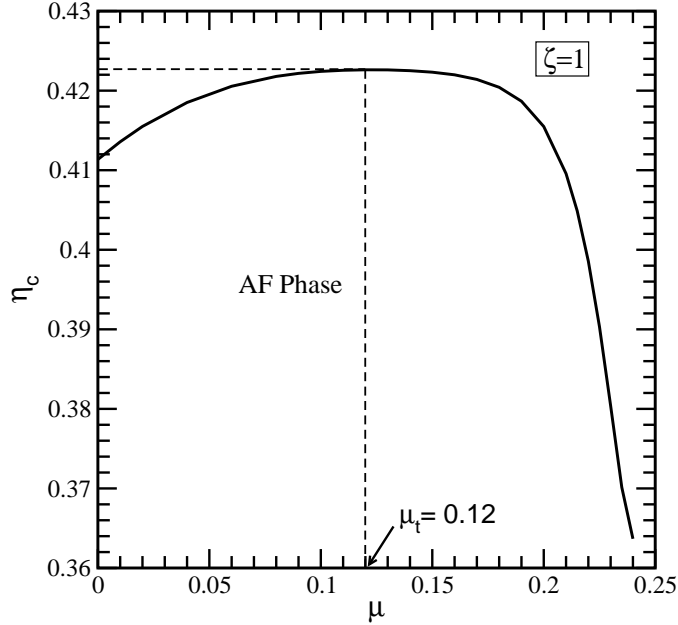}
\caption{(Color online) $\eta_c$-$\mu$ phase diagram for $\zeta=1$. With increase in $\mu$, $\eta_c$ increases up to a maximum value 0.423 at $\mu=\mu_t \approx 0.12$ and then sharply decreases. This 
shows that the ring exchange coupling $\mu$ initially favors the N\'eel 
ordering of the NN spins till the turning value $\mu_t$ is reached. 
For $\mu>\mu_t$, the four-spin coupling enhances destabilizes the N\'{e}el order.}
\label{fig:phasedia1}
\end{figure}

Another interesting feature portrayed in Fig.\ \ref{fig:magdia1} is the change in the critical value of $\eta$ with $\mu$. For $\mu=0$ the magnetization vanishes at the critical value of frustration 
$\eta_c \approx 0.411$. With increase in $\mu$, the value of $\eta_c$ increases initially till
a turning value of $\mu=\mu_t \approx 0.12$ is reached beyond which $\eta_c$ decreases again. 
For example, $\eta_c \approx 0.423$ for $\mu=0.12$, but $\eta_c \approx 0.399$ for $\mu=0.22$. This implies that the four-spin ring exchange interaction favors the N\'{e}el order and thus extends the AF
region only for small values. Beyond the turning value $\mu=\mu_t$ is reached the ring exchange coupling 
destabilizes the N\'eel phase. This is shown in the $\eta_c$-$\mu$ phase diagram in Fig.\ \ref{fig:phasedia1}.

Next we study the influence of directional anisotropy between the horizontal and vertical NN couplings
implying $\zeta < 1$. This spatial anisotropy does not lead to frustration, but it weakens the
NN coupling because the vertical NN coupling is lowered. Hence we expect a qualitatively
similar behavior as before, but at lower values of $\eta$ and $\mu$. This expectation is
confirmed by the following results.

\begin{figure}[httb]%
\centering
\includegraphics[width=3.5in,clip]{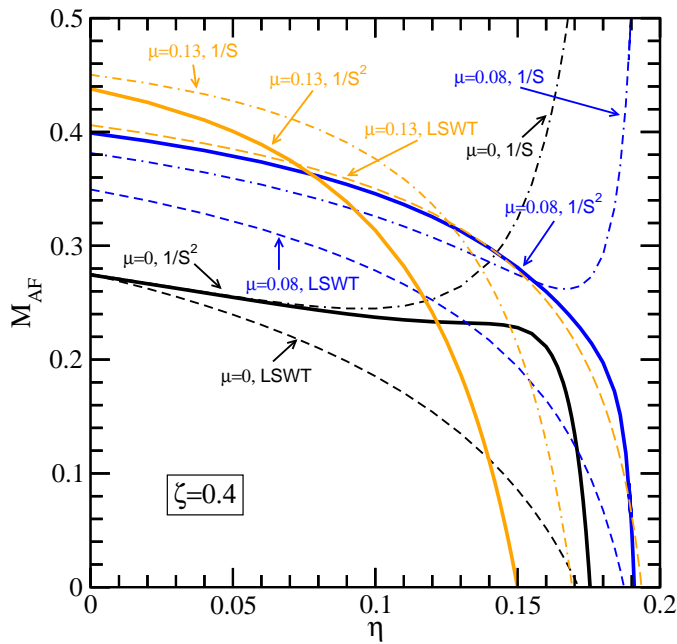}
\caption{(Color online) Sublattice magnetization $M_{\rm AF}$ with spatial anisotropy $\zeta=0.4$ between the vertical and the horizontal NN couplings for three values
of $\mu=0$ (black), 0.08 (blue/dark gray), 0.13 (orange/light gray) as a function of frustration $\eta$. 
For all three cases, results from LSWT (dashed lines), with $1/S$ (dot-dashed lines), and with $1/S^2$ corrections (solid lines) are shown.
$M_{\rm AF}$ with $1/S$ corrections alone diverge for $\mu=0$ and 0.08, but not
for $\mu=0.13$ where it converges, cf.\ main text}.
\label{fig:magdia2}
\end{figure}

Figure \ref{fig:magdia2} shows the magnetization upon increasing $\eta$ for the spatially
 anisotropic case. We choose $\zeta =0.4$ with the three values of ring exchange coupling 
$\mu=0, 0.08$, and 0.13. Here the values of the magnetization without NNN frustration are 
$M_{\rm AF}(\eta=0, \mu=0.08) \approx 0.40$ and  $M_{\rm AF}(\eta=0, \mu=0.13) \approx 0.438$.
Again these numbers are again larger than the LSWT values which are  0.350 and 0.406, respectively.

It is interesting to observe that with increase in $\eta$
the magnetization with just $1/S$ corrections (dotted curves) diverge except for the case when $\mu=0.13$. We find that this divergence ceases to occur for $\mu \gtrapprox 0.10$. 
As before, $1/S^2$ corrections significantly modify the 
magnetization curves. The critical values of $\eta$ at which the N\'eel phase is unstable are 
0.176, 0.191, and 0.15 for $\mu=0, 0.08$ and 0.13, respectively. 
The LSWT predictions for these three cases are 
 0.172, 0.188, and 0.194, respectively. Notice that the LSWT prediction $\eta_c=0.194$ for $\mu=0.13$ 
 is larger than the value $\eta_c=0.15$ obtained including first and second order corrections. 

\begin{figure}[httb]
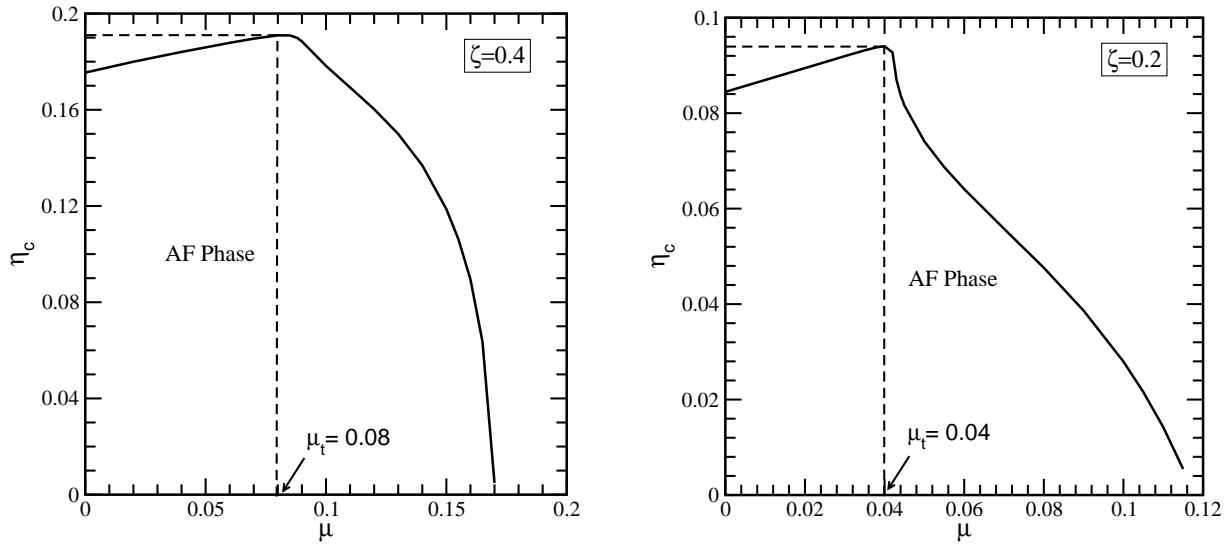

\centering
\includegraphics[width=3.0in,clip]{zeta04-mu-etacphase.eps}
\qquad
\includegraphics[width=3.0in,clip]{zeta02-mu-etacphase.eps}
\caption{(Color online) $\eta_c$-$\mu$ phase diagram for $\zeta=0.4$ (left panel) and 0.2
(right panel), to be compared with the phase diagram for the spatially isotropic case $\zeta=1$ 
in Fig.\ \ref{fig:phasedia1}}.
\label{fig:phasedia2}
\end{figure}

It is worth exploring the influence of the spatial anisotropy 
$\zeta$ on the $\eta_c$-$\mu$ phase diagram. This is done in the panels of Fig.\ \ref{fig:phasedia2} for $\zeta=0.4$ and 0.2. The results are qualitatively similar to those
for $\zeta=1$ in Fig.\ \ref{fig:phasedia1}, but at lower values of $\eta$ and $\mu$
as we expected. The N\'eel phase is stabilized by small values of $\mu$. But beyond
the turning values $\mu_t$ the four-spin ring exchange starts to reduce the 
parameter region of the N\'eel phase.

\section{\label{sec:conclusions}Conclusions}

For $S=1/2$ Heisenberg the four-spin ring exchange coupling on plaquettes
is the next important interaction after the nearest-neighbor exchange.
In this work we have investigated its influence on the zero temperature 
magnetic phase diagram of a spatially anisotropic and frustrated Heisenberg antiferromagnet 
on the square lattice. 

In particular, we studied higher-order quantum effects 
in a systematic perturbative spin-wave expansion in the inverse spin $S$.
We have calculated the spin-wave energy and the magnetization up to and including
the second-order corrections. They contribute significantly to the shape of 
the magnetic phase diagram,  especially as the frustration between the next-nearest neighbor
spins increases. The obtained magnetic phase diagram shows that the four-spin ring exchange coupling initially favors the N\'eel order until a specific turning value is reached.
Beyond this values a further increase in the ring exchange coupling
increases the frustration in the system and reduces the parameter region in which the N\'{e}el order
represents the stable ground state. 

Moreover, we analyzed the available neutron scattering data
and found that a ring exchange coupling $2K$ of about 27\% to 29\% of the
nearest-neighbor exchange is required to explain the data. The additional 
determination of the relative frustration in a three-parameter fit
is not possible because the dispersions for various triples of nearest-neighbor
exchange, frustration, and four-spin ring exchange are indistinguishable
if the energies at $(\pi,0)$ and $(\pi/2,\pi/2)$ are matched.

\section{Acknowledgment}
We are grateful to R. Coldea and S. Hayden for providing the inelastic neutron scattering data.
The authors acknowledge the Texas Advanced Computing Center (TACC) at The University of Texas at Austin for providing HPC resources that have contributed to the research results reported within this paper. 

\appendix
\section{\label{Greensfunction} Green's functions and Self-energies}

\begin{figure}[httb]
\centering
\includegraphics[width=4.5in,clip]{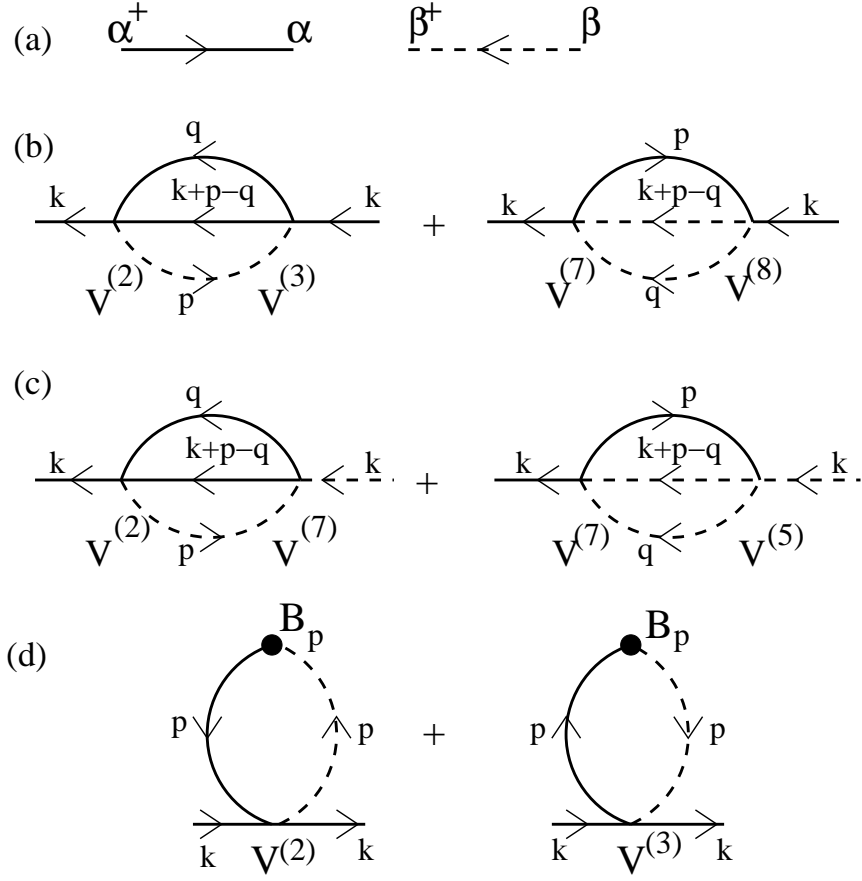}
\caption{\label{fig:Feyn} (a) The solid and the dashed lines correspond to the
$\alpha$ and $\beta$ propagators. Second-order diagrams for the self-energies 
$\Sigma^{(2)}_{\alpha \alpha}({\bf k},\omega)$ and $\Sigma^{(2)}_{\alpha \beta}({\bf k},\omega)$
are shown in (b) and (c). The diagrams in (d) contribute only to 
$\Sigma^{(2)}_{\alpha \alpha}({\bf k},\omega)$. 
$V^{(2)}, V^{(3)}, V^{(5)}, V^{(7)}, V^{(8)}$ are the vertex factors, see main text.
Note that at each vertex two arrows enter the vertex and two leave it which
reflects the conservation of the total $S_z$ component.}
\end{figure}

The time-ordered magnon Green's functions are defined as
\begin{subequations}
\begin{eqnarray}
G_{\alpha \alpha} ({\bf k},t) &=& -i\langle T(\alpha_{\bf k}(t)\alpha^\dag_{\bf k}(0))\rangle,\quad
G_{\beta \beta} ({\bf k},t) = -i\langle T(\beta^\dag_{-\bf k}(t)\beta_{-\bf k}(0))\rangle,
\\
G_{\alpha \beta} ({\bf k},t) &=& -i\langle T(\alpha_{\bf k}(t)\beta_{-\bf k}(0))\rangle,\quad
G_{\beta \alpha} ({\bf k},t) = -i\langle T(\beta^\dag_{-\bf k}(t)\alpha^\dag_{\bf k}(0))\rangle, 
\end{eqnarray}
\end{subequations}
Considering $H_0$ as the unperturbed Hamiltonian the Fourier transformed 
unperturbed propagators are 
\begin{subequations}
\bea
G^0_{\alpha \alpha} ({\bf k},\omega) &=& \frac 1{\omega - E_k+i\delta},\quad
G^0_{\beta \beta} ({\bf k},\omega) = \frac 1{-\omega - E_k+i\delta}, \\
G^0_{\alpha \beta} ({\bf k},\omega) &=& G^0_{\beta \alpha}({\bf k},\omega)=0,
\eea
\end{subequations}
with $\delta \rightarrow 0+$. The spin-wave energy 
$E_{\bf k}=\kappa_{\bf k}\epsilon_{\bf k}$ is measured in units of $J_1S z(1+\zeta-8\mu)$.
The graphical representations of the Green functions are shown in Fig.\ \ref{fig:Feyn}(a).
Note the differing convention for the arrows which help to represent the
conservation of the total $S_z$ component in the diagrams efficiently, see Fig.\ \ref{fig:Feyn}.

The full propagators $G_{ij}({\bf k},\omega)$ satisfy the matrix Dyson equation
\be
G_{ij}({\bf k},\omega)=G^{0}_{ij}({\bf k},\omega)+\sum_{mn}G^0_{im}({\bf k},\omega)
\Sigma_{mn}({\bf k},\omega)G_{nj}({\bf k},\omega),
\ee
where the self-energy $\Sigma_{ij}({\bf k})$ can be expressed in
powers of $1/(2S)$ as 
\be
\Sigma_{ij}({\bf k},\omega)=\frac 1{(2S)}\Sigma_{ij}^{(1)}({\bf k},\omega)
+\frac 1{(2S)^2}\Sigma_{ij}^{(2)}({\bf k},\omega) + \ldots .
\ee
The first-order self-energy terms read
\begin{subequations}
\bea
\Sigma_{\alpha \alpha}^{(1)}({\bf k}, \omega) &=& \Sigma_{\beta \beta}^{(1)}({\bf k}, \omega)=A_{\bf k}, \\
\Sigma_{\alpha \beta}^{(1)}({\bf k}, \omega)&=& \Sigma_{\beta \alpha}^{(1)}({\bf k}, \omega)=B_{\bf k}.
\eea
\end{subequations}

The second-order self-energy terms originate from the Feynman diagrams in Figs.~\ref{fig:Feyn}(b), 
(c), and (d). The coefficients ${\cal C}_{1{\bf k}}$ and ${\cal C}_{2{\bf k}}$ stem
 from the normal-ordering of ${\cal H}_2$. The complete expressions read
 \begin{subequations}
\bea
\Sigma_{\alpha \alpha}^{(2)}({\bf k}, \omega) &=& \Sigma_{\beta \beta}^{(2)}({\bf k}, \omega)={\cal C}_{1{\bf k}}
+ \Big(\frac 2{N}\Big)\sum_{\bf p}(\ell_{\bf k}\ell_{\bf p})^2\frac {B_{\bf p}(V^{(2)}_{\bf k,p,p,k}+
V^{(3)}_{\bf k,p,p,k})}{2E_{\bf p}} \non \\
&+&\Big(\frac 2{N}\Big)^2\sum_{\bf p q}2(\ell_{\bf p}\ell_{\bf q}\ell_{\bf k}\ell_{\bf [k+p-q]})^2
\Big[\frac {V^{(2)}_{\bf k,p,q,[k+p-q]}V^{(3)}_{\bf [k+p-q],q,p,k}}{\omega -E_{\bf p}-E_{\bf q}-E_{\bf [k+p-q]}+i\delta} \non \\
&-& \frac {V^{(7)}_{\bf k,p,q,[k+p-q]}V^{(8)}_{\bf [k+p-q],q,p,k}}{\omega +E_{\bf p}+E_{\bf q}+E_{\bf [k+p-q]}-i\delta}\Big], \label{SEeqs1} \\
\Sigma_{\alpha \beta}^{(2)}({\bf k}, \omega) &=& \Sigma_{\beta \alpha}^{(2)}({\bf k}, \omega)={\cal C}_{2{\bf k}}
+ \Big(\frac 2{N}\Big)^2\sum_{\bf p q}2(\ell_{\bf p}\ell_{\bf q}\ell_{\bf k}\ell_{\bf [k+p-q]})^2\times \non \\
&&\Big[\frac {V^{(2)}_{\bf k,p,q,[k+p-q]}V^{(7)}_{\bf [k+p-q],q,p,k}}{\omega -E_{\bf p}-E_{\bf q}-E_{\bf [k+p-q]}+i\delta}- \frac {V^{(7)}_{\bf k,p,q,[k+p-q]}V^{(5)}_{\bf [k+p-q],q,p,k}}{\omega +E_{\bf p}+E_{\bf q}+E_{\bf [k+p-q]}-i\delta}\Big],
\label{SEeqs2}
\eea
\end{subequations}
where $[{\bf k+p-q}]$ is meant to be mapped to $({\bf k+p-q})$ in the first BZ by an appropriate
reciprocal  vector ${\bf G}$. In deriving Eqs.\  \eqref{SEeqs1} and \eqref{SEeqs2} we have used the symmetry properties of the vertices, see Eq.\ \eqref{vertexsymm}.

\section{\label{VertexAF} Vertex factors}

The expressions for the vertex factors are very lengthy. It is convenient to first define the following functions
\bea
{\cal J}_1 &=& \gamma_2(1-4)+\gamma_2(2-4)+\gamma_2(1-3)+\gamma_2(2-3)-\gamma_2(1)-\gamma_2(2)
-\gamma_2(1-3-4) \non \\
&-& \gamma_2(2-3-4),\non  \\
{\cal J}_2 &=& \gamma_2(1-4)+\gamma_2(2-4)+\gamma_2(1-3)+\gamma_2(2-3), \non\\
{\cal S}_1 &=& \gamma_x(4)\gamma_y(2-4)+\gamma_x(1+2-4)\gamma_y(1-4)+\gamma_x(1-3)\gamma_y(1+2-3)
+\gamma_x(2-3)\gamma_y(3) \non \\
&+&\gamma_x(3)\gamma_y(2-3)+\gamma_x(1+2-3)\gamma_y(1-3)+\gamma_x(1-4)\gamma_y(1+2-4)
+\gamma_x(2-4)\gamma_y(4), \non \\
{\cal S}_2 &=& \gamma_x(4)\gamma_y(1-4)+\gamma_x(1+2-4)\gamma_y(2-4)+\gamma_x(2-3)\gamma_y(1+2-3)
+\gamma_x(1-3)\gamma_y(3) \non \\
&+&\gamma_x(3)\gamma_y(1-3)+\gamma_x(1+2-3)\gamma_y(2-3)+\gamma_x(2-4)\gamma_y(1+2-4)
+\gamma_x(1-4)\gamma_y(4), \non \\
{\cal S}_3 &=& \gamma_x(1-3-4)\gamma_y(2-4)+\gamma_x(1)\gamma_y(2-3)+\gamma_x(1-4)\gamma_y(2-3-4)
+\gamma_x(1-3)\gamma_y(2) \non \\
&+&\gamma_x(2-3-4)\gamma_y(1-4)+\gamma_x(2)\gamma_y(1-3)+\gamma_x(2-4)\gamma_y(1-3-4)
+\gamma_x(2-3)\gamma_y(1), \non \\
{\cal S}_4 &=& \gamma_x(1-3-4)\gamma_y(2-3)+\gamma_x(1)\gamma_y(2-4)+\gamma_x(1-3)\gamma_y(2-3-4)
+\gamma_x(1-4)\gamma_y(2) \non \\
&+&\gamma_x(2-3-4)\gamma_y(1-3)+\gamma_x(2)\gamma_y(1-4)+\gamma_x(2-3)\gamma_y(1-3-4)
+\gamma_x(2-4)\gamma_y(1), \non \\
{\cal S}_5 &=& \gamma_x(2)\gamma_y(2-3)+\gamma_x(2-3-4)\gamma_y(2-4)+\gamma_x(1-4)\gamma_y(1-3-4)
+\gamma_x(1-3)\gamma_y(1) \non \\
&+&\gamma_x(1)\gamma_y(1-3)+\gamma_x(1-3-4)\gamma_y(1-4)+\gamma_x(2-4)\gamma_y(2-3-4)
+\gamma_x(2-3)\gamma_y(2), \non \\
{\cal S}_6 &=& \gamma_x(2)\gamma_y(2-4)+\gamma_x(2-3-4)\gamma_y(2-3)+\gamma_x(1-3)\gamma_y(1-3-4)
+\gamma_x(1-4)\gamma_y(1) \non \\
&+&\gamma_x(1)\gamma_y(1-4)+\gamma_x(1-3-4)\gamma_y(1-3)+\gamma_x(2-3)\gamma_y(2-3-4)
+\gamma_x(2-4)\gamma_y(2), \non \\
{\cal S}_7 &=& \gamma_x(1+2-3)\gamma_y(1-4)+\gamma_x(3)\gamma_y(2-4)+\gamma_x(1-3)\gamma_y(1+2-4)
+\gamma_x(2-3)\gamma_y(4) \non \\
&+&\gamma_x(1+2-4)\gamma_y(1-3)+\gamma_x(4)\gamma_y(2-3)+\gamma_x(1-4)\gamma_y(1+2-3)
+\gamma_x(2-4)\gamma_y(3), \non \\
{\cal S}_8 &=& \gamma_x(1+2-3)\gamma_y(2-4)+\gamma_x(3)\gamma_y(1-4)+\gamma_x(2-3)\gamma_y(1+2-4)
+\gamma_x(1-3)\gamma_y(4) \non \\
&+&\gamma_x(1+2-4)\gamma_y(2-3)+\gamma_x(4)\gamma_y(1-3)+\gamma_x(2-4)\gamma_y(1+2-3)
+\gamma_x(1-4)\gamma_y(3), \non \\
{\cal S}_9 &=& \gamma_x(1-4)\gamma_y(2-4)+\gamma_x(1-3)\gamma_y(2-3)+\gamma_x(2-4)\gamma_y(1-4)
+\gamma_x(2-3)\gamma_y(1-3), \non \\
{\cal S}_{10} &=& \gamma_x(2-3)\gamma_y(2-4)+\gamma_x(1-3)\gamma_y(1-4)+\gamma_x(2-4)\gamma_y(2-3)
+\gamma_x(1-4)\gamma_y(1-3), \non \\
{\cal S}_{11} &=& \gamma_x(2)\gamma_y(4)+\gamma_x(4)\gamma_y(2)+\gamma_x(1+2-3)\gamma_y(1-3-4)
+\gamma_x(1-3-4)\gamma_y(1+2-3) \non \\
&+&\gamma_x(1+2-4)\gamma_y(1)+\gamma_x(1)\gamma_y(1+2-4)+\gamma_x(2-3-4)\gamma_y(3)
+\gamma_x(3)\gamma_y(2-3-4), \non \\
{\cal S}_{12} &=& \gamma_x(2)\gamma_y(3)+\gamma_x(3)\gamma_y(2)+\gamma_x(1+2-4)\gamma_y(1-3-4)
+\gamma_x(1-3-4)\gamma_y(1+2-4) \non \\
&+&\gamma_x(1+2-3)\gamma_y(1)+\gamma_x(1)\gamma_y(1+2-3)+\gamma_x(2-3-4)\gamma_y(4)
+\gamma_x(4)\gamma_y(2-3-4), \non 
\eea
\bea
{\cal S}_{13} &=& \gamma_x(1)\gamma_y(4)+\gamma_x(4)\gamma_y(1)+\gamma_x(1+2-3)\gamma_y(2-3-4)
+\gamma_x(2-3-4)\gamma_y(1+2-3) \non \\
&+&\gamma_x(1+2-4)\gamma_y(2)+\gamma_x(2)\gamma_y(1+2-4)+\gamma_x(1-3-4)\gamma_y(3)
+\gamma_x(3)\gamma_y(1-3-4), \non \\
{\cal S}_{14} &=& \gamma_x(1)\gamma_y(3)+\gamma_x(3)\gamma_y(1)+\gamma_x(1+2-4)\gamma_y(2-3-4)
+\gamma_x(2-3-4)\gamma_y(1+2-4) \non \\
&+&\gamma_x(1+2-3)\gamma_y(2)+\gamma_x(2)\gamma_y(1+2-3)+\gamma_x(1-3-4)\gamma_y(4)
+\gamma_x(4)\gamma_y(1-3-4). \non 
\eea

The vertex factors required for our calculations are
\begin{subequations}
\bea
V^{(2)}_{12;34} &=& \Big[-x_3\gamma_1(2-3)-x_4\gamma_1(2-4)-x_1x_2x_3\gamma_1(1-3)
-x_1x_2x_4 \gamma_1(1-4) \non \\
&+& x_1x_2\gamma_1(1)+\gamma_1(2)+x_1x_2x_3x_4\gamma_1(1-3-4)
+x_3x_4\gamma_1(2-3-4)\Big ] \non \\
&+&\Big(\frac {\eta-2\mu}{1+\zeta-8\mu}\Big)\Big[x_2+\Phi_G x_1x_3x_4\Big]{\cal J}_1  \non \\
&-&\Big(\frac {4\mu}{1+\zeta-8\mu}\Big)\Big[ -(x_2+\Phi_G x_3x_4){\cal J}_2
+\half ({\cal S}_1+x_1x_2{\cal S}_2+x_1x_3{\cal S}_3 
+ x_1x_4{\cal S}_4 \non \\
&+&x_2x_3{\cal S}_5 + x_2x_4{\cal S}_6 
+ x_3x_4{\cal S}_7+x_1x_2x_3x_4{\cal S}_8 - 2x_1{\cal S}_9-2x_2x_3x_4{\cal S}_{10}
-x_4{\cal S}_{11}\non \\ 
&-& x_3{\cal S}_{12} - x_1x_2x_4{\cal S}_{13}-x_1x_2x_3{\cal S}_{14})\Big].
\\
V^{(3)}_{12;34} &=& \Big[-x_1\gamma_1(1-3)-x_2\gamma_1(2-3)-x_1x_3x_4\gamma_1(1-4)
-x_2x_3x_4 \gamma_1(2-4) \non \\
&+& x_1x_3\gamma_1(1)+x_2x_3\gamma_1(2)+x_1x_4\gamma_1(1-3-4)
+x_2x_4\gamma_1(2-3-4)\Big ] \non \\
&+&\Big(\frac {\eta-2\mu}{1+\zeta-8\mu}\Big)\Big[x_3+\Phi_G x_1x_2x_4\Big]{\cal J}_1  \non \\
&-&\Big(\frac {4\mu}{1+\zeta-8\mu}\Big)\Big[ -(x_3+\Phi_G x_1x_2x_4){\cal J}_2
+\half (x_2x_3{\cal S}_1+x_1x_3{\cal S}_2+x_1x_2{\cal S}_3 
+ x_1x_2x_3x_4{\cal S}_4 \non \\
&+&{\cal S}_5 + x_3x_4{\cal S}_6 
+ x_2x_4{\cal S}_7+x_1x_4{\cal S}_8 - 2x_1x_2x_3{\cal S}_9-2x_4{\cal S}_{10}
-x_2x_3x_4{\cal S}_{11}\non \\ 
&-& x_2{\cal S}_{12} - x_1x_3x_4{\cal S}_{13}-x_1{\cal S}_{14})\Big].
\\
V^{(5)}_{12;34} &=& \Big[-x_2x_3x_4\gamma_1(1-3)-x_1x_3x_4\gamma_1(2-3)-x_1\gamma_1(2-4)
-x_2\gamma_1(1-4) \non \\
&+& x_1x_4\gamma_1(2)+x_2x_4\gamma_1(1)+x_1x_3\gamma_1(2-3-4)
+x_2x_3\gamma_1(1-3-4)\Big ] \non \\
&+&\Big(\frac {\eta-2\mu}{1+\zeta-8\mu}\Big)\Big[x_1x_2x_4+\Phi_G x_3\Big]{\cal J}_1  \non \\
&-&\Big(\frac {4\mu}{1+\zeta-8\mu}\Big)\Big[ -(x_1x_2x_4+\Phi_G x_3){\cal J}_2
+\half (x_1x_4{\cal S}_1+x_2x_4{\cal S}_2+x_3x_4{\cal S}_3 
+ {\cal S}_4\non \\ 
&+& x_1x_2x_3x_4{\cal S}_5 + x_1x_2{\cal S}_6 
+ x_1x_3{\cal S}_7+x_2x_3{\cal S}_8 - 2x_4{\cal S}_9-2x_1x_2x_3{\cal S}_{10}
-x_1{\cal S}_{11}\non \\ 
&-& x_1x_3x_4{\cal S}_{12} - x_2{\cal S}_{13}-x_2x_3x_4{\cal S}_{14})\Big].
\eea

\bea
V^{(7)}_{12;34} &=& \Big[x_1x_4\gamma_1(1-3)+x_1x_3\gamma_1(1-4)+x_2x_3\gamma_1(2-4)
+x_2x_4\gamma_1(2-3) \non \\
&-& x_1x_3x_4\gamma_1(1)-x_2x_3x_4\gamma_1(2)-x_1\gamma_1(1-3-4)
-x_2\gamma_1(2-3-4)\Big ] \non \\
&+&\Big(\frac {\eta-2\mu}{1+\zeta-8\mu}\Big)\Big[-x_3x_4-\Phi_G x_1x_2\Big]{\cal J}_1  \non \\
&-&\Big(\frac {4\mu}{1+\zeta-8\mu}\Big)\Big[ (x_3x_4+\Phi_G x_1x_2){\cal J}_2
+\half (-x_2x_3x_4{\cal S}_1-x_1x_3x_4{\cal S}_2-x_1x_2x_4{\cal S}_3 \non \\
&-& x_1x_2x_3 {\cal S}_4- x_4{\cal S}_5 - x_3{\cal S}_6 
- x_2{\cal S}_7-x_1{\cal S}_8 +2x_1x_2x_3x_4{\cal S}_9+2{\cal S}_{10}
+x_2x_3{\cal S}_{11}\non \\ 
&+& x_2x_4{\cal S}_{12} + x_1x_3{\cal S}_{13}+x_1x_4{\cal S}_{14})\Big].
\\
V^{(8)}_{12;34} &=& \Big[x_1x_4\gamma_1(2-4)+x_2x_4\gamma_1(1-4)+x_1x_3\gamma_1(2-3)
+x_2x_3\gamma_1(1-3) \non \\
&-& x_1\gamma_1(2)-x_2\gamma_1(1)-x_1x_3x_4\gamma_1(2-3-4)
-x_2x_3x_4\gamma_1(1-3-4)\Big ] \non \\
&+&\Big(\frac {\eta-2\mu}{1+\zeta-8\mu}\Big)\Big[-x_1x_2-\Phi_G x_3x_4\Big]{\cal J}_1  \non \\
&-&\Big(\frac {4\mu}{1+\zeta-8\mu}\Big)\Big[ (x_1x_2+\Phi_G x_3x_4{)\cal J}_2
+\half (-x_1{\cal S}_1-x_2{\cal S}_2-x_3{\cal S}_3- x_4 {\cal S}_4\non \\
&-& x_1x_2x_3{\cal S}_5- x_1x_2x_4{\cal S}_6 
- x_1x_3x_4{\cal S}_7-x_2x_3x_4{\cal S}_8 +2{\cal S}_9+2x_1x_2x_3x_4{\cal S}_{10}
\non \\ 
&+& x_1x_4{\cal S}_{11}+x_1x_3{\cal S}_{12} + x_2x_4{\cal S}_{13}+x_2x_3{\cal S}_{14})\Big],
\eea
\end{subequations}
where $\Phi_G=\exp(iG_x)$, $G_x$ being the $x$-component of the reciprocal lattice vector ${\bf G}$
appearing in the momentum conserving delta-function in Eq.\ \eqref{H1term}.
These vertex factors fulfill the following symmetry relations
\begin{subequations}
\label{vertexsymm}
\begin{eqnarray}
V^{(2)}_{12;34} &=& V^{(2)}_{12;43};\quad
V^{(3)}_{12;34}=V^{(3)}_{21;34};\quad
V^{(5)}_{12;34}=V^{(5)}_{21;34},  \\
V^{(7)}_{12;34} &=& V^{(7)}_{21;34}=V^{(7)}_{12;43};\quad
 V^{(8)}_{12;34}=V^{(8)}_{21;34}=V^{(8)}_{12;43}.
\end{eqnarray}
If no reciprocal lattice vector is involved in the momentum conservation, i.e., ${\bf G}=0$,
 there are some additional symmetries
\begin{equation}
V^{(3)}_{12;34}=V^{(5)}_{12;34};\quad
V^{(7)}_{12;34} = V^{(8)}_{12;34}.
\end{equation}
\end{subequations}

\section{\label{Ck} Coefficients ${\cal C}_{1{\bf k}}$ and ${\cal C}_{2{\bf k}}$}

We define the functions ${\cal P}_{\bf k}$ and ${\cal Q}_{\bf k}$
\begin{subequations}
\bea
{\cal P}_{\bf k} &=& \Big(\frac 2{N} \Big)^2\sum_{12} 2\ell_1^2 \ell_2^2\Big[
x_1^2x_2^2\Big\{6+6\gamma_2(k)+6\gamma_2(2)+2\gamma_2(k-2) +\gamma_x(1-2)\gamma_y(1+2)\non \\
&+&\gamma_x(k-1-2)\gamma_y(k-1+2)\Big\}+x_1^2\Big\{2\gamma_2(k-2)+6\gamma_2(2)+\gamma_x(k+1-2)\gamma_y(k-1-2) \non \\
&+& \gamma_x(k-1-2)\gamma_y(k+1-2)+\gamma_x(k-1+2)\gamma_y(k-1-2)+\gamma_x(k-1+2)\gamma_y(k+1-2)\Big\}\non \\
&+& x_1x_2\Big\{ 4\gamma_x(k-1)\gamma_y(k-2)+4\gamma_x(k-2)\gamma_y(k-1)+4\gamma_x(1)\gamma_y(2)
+4\gamma_x(2)\gamma_y(1)\non \\
&+& 6\gamma_x(1-2)+6\gamma_y(1-2)+4\gamma_x(k)\gamma_y(k-1-2)+4\gamma_x(k-1-2)\gamma_y(k)\Big\}\non \\
&-& x_1^2x_2\Big\{ 8\gamma_x(k)\gamma_y(k-2)+8\gamma_x(k-2)\gamma_y(k)+12\gamma_x(2)+12\gamma_y(2)
+4\gamma_x(1)\gamma_y(1-2)+ \non \\
&+& 4\gamma_x(1-2)\gamma_y(1)+2\gamma_x(k-1)\gamma_y(k-1-2)+2\gamma_x(k-1-2)\gamma_y(k-1) \non \\
&+& 2\gamma_x(k+1-2)\gamma_y(k-1)+2\gamma_x(k-1)\gamma_y(k+1-2)\Big\}\non \\
&-&x_1\Big\{4\gamma_x(2)\gamma_y(1-2)+4\gamma_x(1-2)\gamma_y(2)+2\gamma_x(k-2)\gamma_y(k-1+2)
\non \\
&+&2\gamma_x(k-1+2)\gamma_y(k-2)+ 2\gamma_x(k+1-2)\gamma_y(k-2)+2\gamma_x(k-2)\gamma_y(k+1-2) \Big\}\non \\
&+&\Big\{ \gamma_x(1-2)\gamma_y(1+2)+\gamma_x(k-1-2)\gamma_y(k+1-2)\Big\}
\Big].
\\
{\cal Q}_{\bf k} &=& \Big(\frac 2{N} \Big)^2\sum_{12}2 \ell_1^2 \ell_2^2\Big[
x_1^2x_2^2\Big\{6\gamma_x(k)+6\gamma_y(k)+4\gamma_x(2)\gamma_y(k-2) +4\gamma_x(k-2)\gamma_y(2)\non \\
&+&\gamma_x(k-1-2)\gamma_y(1-2)+\gamma_x(1-2)\gamma_y(k-1-2)\Big\} \non \\
&+& x_1^2\Big\{4\gamma_x(2)\gamma_y(k-2)+4\gamma_x(k-2)\gamma_y(2) 
+ \gamma_x(k+1-2)\gamma_y(1-2)\non \\
&+& \gamma_x(1-2)\gamma_y(k+1-2)+\gamma_x(k-1+2)\gamma_y(1-2)
+\gamma_x(1-2)\gamma_y(k-1+2)\Big\}\non \\
&+& x_1x_2\Big\{ 8\gamma_x(2)\gamma_y(k-1)+8\gamma_x(k-1)\gamma_y(2)+6\gamma_x(k-1+2)+6\gamma_y(k-1+2)
\non \\
&+& 4\gamma_x(k)\gamma_y(1-2)+4\gamma_x(1-2)\gamma_y(k)\Big\}\non \\
&-& x_1^2x_2\Big\{ 8\gamma_x(k)\gamma_y(2)+8\gamma_x(2)\gamma_y(k)+12\gamma_x(k-2)+12\gamma_y(k-2)
+4\gamma_x(k-1)\gamma_y(1-2)+ \non \\
&+& 4\gamma_x(1-2)\gamma_y(k-1)+4\gamma_x(k-1-2)\gamma_y(1)+4\gamma_x(1)\gamma_y(k-1-2)\Big\}\non \\
&-&x_1\Big\{4\gamma_x(k-2)\gamma_y(1-2)+4\gamma_x(1-2)\gamma_y(k-2)+4\gamma_x(2)\gamma_y(k-1-2)\non \\
&+& 4\gamma_x(k-1-2)\gamma_y(2)\Big\} 
+\Big\{ \gamma_x(1-2)\gamma_y(k-1-2)+\gamma_x(k-1-2)\gamma_y(1-2)\Big\}
\Big].
\eea
\end{subequations}
Then, the static second-order corrections are given by
\begin{subequations}
\label{Ceqn}
\bea
{\cal C}_{1{\bf k}} &=& (\ell_{\bf k}^2+m_{\bf k}^2){\cal Q}_{\bf k}+2\ell_{\bf k}m_{\bf k}{\cal P}_{\bf k},
\\
{\cal C}_{2{\bf k}} &=& (\ell_{\bf k}^2+m_{\bf k}^2){\cal P}_{\bf k}+2\ell_{\bf k}m_{\bf k}{\cal Q}_{\bf k}.
\eea
\end{subequations}

\bibliography{Cyclic}

\end{document}